# Open-Source Concealed EEG Data Collection for Brain-Computer-Interfaces

Open-Source Concealed Ear-EEG

Real-World Neural Observation Through OpenBCI Amplifiers with Around-the-Ear cEEGrid Electrodes

Michael Thomas Knierim

Institute of Information Systems and Marketing (IISM), Karlsruhe Institute of Technology (KIT), Karlsruhe, Germany, michael.knierim@kit.edu

Christoph Berger

Institute of Information Systems and Marketing (IISM), Karlsruhe Institute of Technology (KIT), Karlsruhe, Germany, christoph.berger9@kit.edu

Pierluigi Reali

Department of Electronics, Information, and Bioengineering, Politecnico di Milano, Milan, Italy, pierluigi.reali@polimi.it

Observing brain activity in real-world settings offers exciting possibilities like the support of physical health, mental well-being, and thought-controlled interaction modalities. The development of such applications is, however, strongly impeded by poor accessibility to research-grade neural data and by a lack of easy-to-use and comfortable sensors. This work presents the cost-effective adaptation of concealed around-the-ear EEG electrodes (cEEGrids) to the open-source OpenBCI EEG signal acquisition platform to provide a promising new toolkit. An integrated system design is described, that combines publicly available electronics components with newly designed 3D-printed parts to form an easily replicable, versatile, single-unit around-the-ear EEG recording system for prolonged use and easy application development. To demonstrate the system's feasibility, observations of experimentally induced changes in visual stimulation and mental workload are presented. Lastly, as there have been no applications of the cEEGrids to HCI contexts, a novel application area for the system is investigated, namely the observation of flow experiences through observation of temporal Alpha power changes. Support for a link between temporal Alpha power and flow is found, which indicates an efficient engagement of verbal-analytic reasoning with intensified flow experiences, and specifically intensified task absorption.

# 1 INTRODUCTION

Observing brain activity in real-world settings offers exciting possibilities like the support of physical health, mental well-being, and thought-controlled human-computer-interaction (HCI) modalities [10,43]. At present, especially the electroencephalogram (EEG), which records the electrical discharges of millions of neurons, is used to realize these goals due to its portability and high temporal resolution – two unparalleled factors among neuroimaging methods [20]. EEG-based Brain-Computer-Interfaces (BCI) are vividly researched in the form of either passive observation systems or active input systems [10,43]. Passive BCI applications approach, for example, the inference of sleep quality [11], the monitoring of mental workload and stress dynamics [45,69], or the detection of dangerous attention lapses in traffic [2]. Active BCI systems leverage intentionally generated brain activity as a user interface to control digital and robotic applications. These systems are, for instance, used with paralyzed patients to re-enable communication with the outside world through speller interfaces [10], but also in the entertainment industry to deliver enticing experiences of mind-controlled gaming or artistry [1,58]. Altogether, BCI scholars envision a reality in which neural changes are continuously monitored to overcome ailments and physical limitations, improve our overall quality of life, and change how humans interface with technology [11,13,43].

The development of such EEG BCI applications is, however, strongly impeded by poor accessibility to research-grade neural data and by a lack of easy-to-use and comfortable sensors [13,22]. The low accessibility typically stems either from the high price point of sensor systems or the complicated assembly requirements for newly engineered system prototypes in scientific publications [65]. The low comfort stems from either having to use gel-based cap systems that require excessive cleaning of skin and hair or from high mechanical pressure on the head to enable good electrode-skin contact [13,22]. Yet, the repeated and continuous acquisition of clean neural signals is the cornerstone for BCIs [10,18]. For these reasons, in recent years, increasing efforts have been put forward to realize low-cost, portable, and high-quality brain activity recordings. At the forefront of this development are open-source EEG signal acquisition systems [65] and unobtrusive EEG electrodes placed inside or around the ear [13,15]. An exemplary recording system that delivers on the aspiration of reducing recording costs and enabling research growth is the open-source OpenBCI platform[1] [65]. The OpenBCI system has already been employed in numerous studies and has demonstrated viable recording quality [65]. A major limitation of this system is, however, that the standard headgear is large and visually prominent (see Figure 1), which limits user acceptance in everyday life [13,22]. In contrast, ear-EEG systems offer much less obvious electrode placement (see Figure 1), while maintaining the benefits of easy application, high comfort and signal quality [13,22]. An electrode type that appears especially promising is the flexible and printed c-shaped EEG electrode array (cEEGrid) placed around the ear. In comparison to in-ear-EEG systems, the advantages of cEEGrids lie in the availability of multiple electrodes that are spaced further apart, which improved the detection of generators of electrical neural activity [11,47]. Furthermore, the cEEGrid electrodes are readily purchasable and reusable without additional personalization or engineering efforts. So far, the OpenBCI platform's benefits have not been joined with the benefits of the cEEGrids. Therefore, cEEGrids remain usable only with other, far more expensive research-grade amplifiers.

---

[1] https://openbci.com/

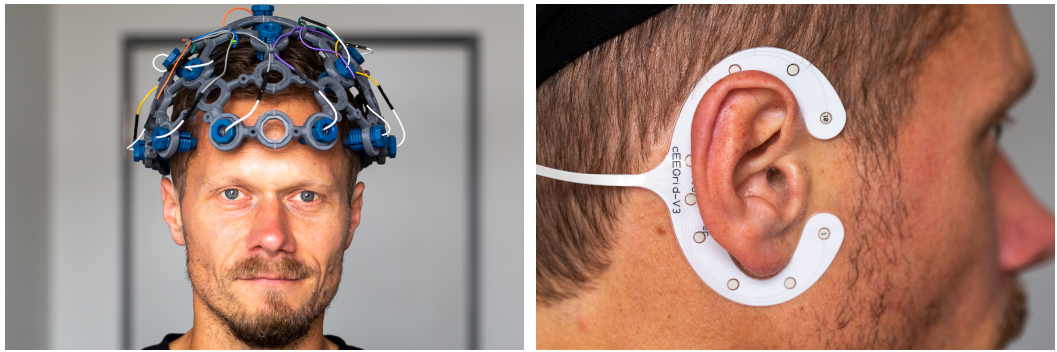

Figure 1: Left: The OpenBCI Ultracortex Mark IV EEG headset. Right: The around-the-ear EEG electrode array (cEEGrid).

To overcome this hurdle, this work evaluates the usability of the OpenBCI platform with cEEGrids. As a basis for the work, the following chapters detail the present state of research on ear-EEG and related OpenBCI extensions, together with a description of the still rather young research on cEEGrids. Afterwards, an integrated system design is described, that combines publicly available electronics components with newly designed 3D-printed parts to form a coherent, single-unit around-the-ear EEG recording system. Detailed instructions and materials for the replication of the system are provided. Following, the technical feasibility of this OpenBCI-cEEGrid extension is evaluated. This technical evaluation represents an essential contribution as the cEEGrids present a challenging setup for EEG amplifiers (e.g. sufficient sensitivity to the small signal amplitudes that are measurable with cEEGrids is needed - [47]). To demonstrate the system's feasibility, observations of experimentally induced changes in visual stimulation and mental workload are presented and the previously indicated extractability of ECG components is quantified. Lastly, as there have been no applications of the cEEGrids to HCI contexts so far, a novel application area for the system is investigated, namely the observation of flow experiences through the observation of temporal Alpha power changes. Support for a link between temporal Alpha power and flow is found, which indicates an efficient engagement of verbal-analytic reasoning with intensified flow experiences, and specifically intensified task absorption. Based on these results, the benefits of the system and future application possibilities are discussed. The major benefits of this work are the integration and evaluation of a system that (1) can easily be employed by BCI and HCI researchers (e.g. without further engineering like electrode assembly or personalization), (2) at a much lower price point (~1.500 USD compared to current alternatives costing ~10.000 USD), (3) with open access to EEG data through versatile application programming interfaces (which greatly facilitates application development), altogether as (4) a wearable solution (which is something that previous OpenBCI ear EEG extensions do not provide).

## 2 RELATED WORK

### 2.1 Ear-Based Sensing Research

Ear-based sensing represents an attractive direction for developers of wearable HCI technologies due to the multitude of detectable signals from within and around the ears (heart and respiratory rates, skin conductance, eye blink and motion signals, and electrical activity from muscles and the brain), and the relative ease and inconspicuousness with which the necessary hardware can be positioned (ear-pieces, headphones, glasses, or other headwear) [13,31,52]. Ear-EEG, in particular, has seen an increased interest as an alternative for otherwise cumbersome or impractical cap-based EEG recordings, as it enables insight into ongoing neural activity indicative of cognitive and affective processes [13,31]. For these reasons, an increasing amount of research has been put forward over the last five years to demonstrate the feasibility of such sensing solutions and to illustrate application scenarios. Given the novelty of ear EEG solutions, a major focus has been on developing first understandings of the signals, compare them to the conventional measurement methods, and explore if the promises of comfort are, in fact, realizable [11]. Such research has for example found the suitability of ear-EEG for continuous daily observation [22], the observation of auditory attention processes [29,35,38,53,59], visual stimulation intensities [22,60], or sleep stage classification [51,56,66]. Due to the advanced understandings, recent work has also started to explore more HCI-related application scenarios like the monitoring of affect [4] and stress [45], biosignal based user authentication [48], and the recognition of mental gestures as user input information for active BCIs [49]. While this research represents a fascinating and growing development in the HCI domain, its progress is greatly impeded by a low degree of accessibility to the required components: (1) cost-effective amplifiers with open data access and (2) ready-to-use ear-EEG electrodes, and (3) the integration of both parts into wearable units that can be used in realistic HCI scenarios. However, two main developments form the basis for this article that provides an opportunity to overcome these limitations: the publicly available OpenBCI biosignal platform and the publicly available cEEGrid electrodes.

## 2.2 The OpenBCI Platform and Previous Ear-EEG Extensions

The OpenBCI platform offers various general-purpose biosignal acquisition devices, featuring an 8 to 16 channel EEG acquisition system for dry or gelled-electrode recordings, with open-source access to all acquired materials and data through public repositories and APIs. The flagship component of the OpenBCI platform is the Cyton board that comprises a microcontroller, amplifier for electrical recordings, wireless connectivity, and SD card recording options for up to 8 recording channels over which electrophysiological data can be collected (brain, heart, and muscles). The system is very affordable in comparison to clinical-grade systems (~1.500 USD for the Cyton + Daisy board) and offers adaptability of hardware and software components through open-source access to all related materials, starting with the 3D-printed electrodes, and mounting harness called the Ultracortex Mark IV (see Figure 1). While the Cyton board contains the essential electronics for the amplifier, the Daisy shield enables data acquisition from eight additional channels. Therefore, through the combination of the Cyton and Daisy boards, this OpenBCI system allows collecting EEG signals with a sampling frequency of 125 Hz (250 Hz or more if the data is recorded to an SD card), data resolution of 24 bits, and compatibility with active and passive electrodes, amongst other features. The validity of the OpenBCI system for the recording of high-quality EEG data has been confirmed in several validation studies in which the 16-channel configuration of the device was compared to clinical-grade EEG acquisition systems from different manufacturers [28,65].

Researchers have used the OpenBCI EEG system to, for instance, observe attention lapses during cycling in traffic [2] or to enable a concentration-based BCI game [3]. Beyond this default system configuration (Cyton + Ultracortex headset), a recent review study on low-cost EEG data acquisition systems highlights the OpenBCI system's adaptability to usage with other components through 3D-printed parts and exchangeable electrodes [65]. Demonstrating this potential, related work has, for example, explored the usage of gold cup electrodes integrated into headbands together with the OpenBCI Cyton board [19]. Concerning ear-EEG recordings, the OpenBCI amplifier has also already been employed with a particular kind of ear-EEG, namely in-ear EEG (where the electrode is placed in the ear canal). This work demonstrates that a combination of the OpenBCI amplifiers with different types of self-developed in-ear electrodes is generally suitable to collect valuable ear-EEG data for purposes such as user authentication [48] or attention monitoring [36,38]. These systems place electrodes either in foam-based or customized earbuds (from individual ear shape molds), approaches that require engineering expertise and are not easily transferable between users due to the need for personalization. Therefore, while this work represents fascinating directions to the HCI community, the technologies still often come with low accessibility. Also, none of these systems tackle the challenge of providing actually wearable designs that would enable naturalistic data acquisition. Presently, these limitations hinder a more extensive investigation of ear-EEG sensing for the HCI community. The herein presented work proposes to close this gap by extending the OpenBCI platform to the use of commercially available around-the-ear EEG electrodes (cEEGrids) and providing a wearable design for the integrated components. This greatly reduces the need for engineering efforts for HCI-scholars interested in EEG sensing, a need that has even been highlighted in the previously mentioned review on OpenBCI extensions [65].

### 2.3 Accessible Around-the-Ear Electrodes: cEEGrids

The cEEGrids are described as a "flexible printed Ag/AgCl electrodes system consisting of ten electrodes arranged in a c-shape to fit around the ear" [23] (see Figure 1 and Figure 2 for a picture and Figure 4 for the layout of the electrodes). This electrode array has been developed to unobtrusively and comfortably collect EEG data in field settings. The cEEGrids have now been repeatedly reported to enable high-quality and multiple hour EEG recordings [15,22,53,60]. The EEG recording quality is primarily realized by the possibility of using the cEEGrids with a gel enclosed by the adhesive and therefore remains fluid for long periods without causing discomfort or requiring the hair to be washed [22]. The electrodes' application around the ear is realized in a few minutes (including light cleaning of the skin with alcohol or an abrasive gel). The electrodes can be re-used numerous times after cleaning the gel residue and re-applying a double-sided adhesive [22]. In comparison to in-ear EEG systems, the advantage of cEEGrids lie in the availability of multiple electrodes that are spaced further apart, which improved the detection of generators of electrical neural activity [11,47], and in their accessibility (they can be readily purchased) and universality (they can be worn without further personalization).

So far, the cEEGrids have demonstrated their ability to record typical EEG phenomena related to visual stimulation (posterior Alpha increases with reduced visual information - see [22]), to auditory stimulation (detection of directed speaker attention – see [29,35,53,59]), sleep stage detection [11,51,66] and changes in mental workload in driving simulations with older adults [69]. These works highlight that the research on cEEGrids has, up to now, primarily focused on answering rather fundamental EEG methodology questions. No dedicated research has been put forward in the HCI context. While the system posits numerous advantages, some challenges limit its acceptance and access. First of all, despite the possibility to use cEEGrids with various amplifiers, most of those used in previous research are fairly costly and might thus not be in reach for researchers, makers, or hackers with a smaller budget. Second of all, more research is still needed to learn about which phenomena can be robustly observed with the cEEGrids in the presence of muscle and movement artefacts [11]. While the latter issues may be reduced by algorithmic means [11], the first hurdle has likely limited the spread of this intriguing new technology.

To increase the accessibility, we considered adapting the cEEGrids to a more affordable biosignal acquisition system, namely the OpenBCI platform. In doing so, we focus not only on providing instructions and materials for the easy replication of the system (see Appendix A.2 and the supplementary materials), but also a technical evaluation. Despite the general possibility of using cEEGrids with various amplifiers [22], this technical evaluation is an essential contribution as the cEEGrids present a challenging setup for EEG amplifiers (sufficient sensitivity to the small signal amplitudes that are measurable with cEEGrids is needed – see [47]). It is not self-evident that cEEGrids would work with just any amplifier, as an amplifier needs to be able to differentiate the neural signals of interest that are tiny compared to other influences (e.g., due to muscle movements). Therefore, through this work, the HCI community gains access to a system that could be easily used by scholars without the need for further engineering, at a much lower price point, with open access to the collected EEG data, and as a wearable solution (which other ear-EEG adaptions to the OpenBCI platform have so far not provided).

## 3 SYSTEM DESCRIPTION

### 3.1 System Component Overview

The herein described system is comprised of three main components (see Figure 2): (1) The OpenBCI Cyton board with the Daisy shield that enables the low-cost mobile biosignal acquisition (e.g., EEG, ECG, or EMG) on 16 recording channels, (2) the cEEGrid electrodes, a set of flexible printed Ag/AgCl electrodes in a c-shaped form that can be placed around a person's ear using double-sided adhesives, and (3) the cEEGrid connector that is comprised of a mini edge card socket that receives the end of the cEEGrid and is soldered on to a printed circuit board (PCB) which transmits the signal from the applied cEEGrids to a biosignal amplifier.

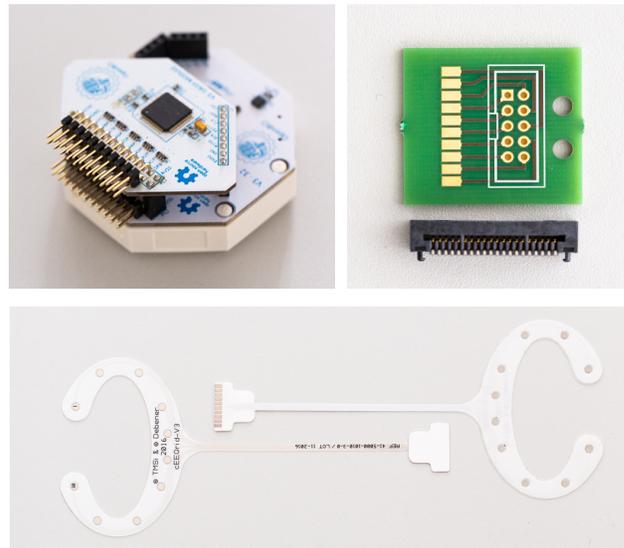

Figure 2: The three components of the herein described system. Top left: OpenBCI Cyton Board and Daisy Shield in the 3D-printed enclosure by OpenBCI. Top right: Printed circuit board (PCB) with mini edge card socket to connect the cEEGrids. Bottom: Outer and inner view of the cEEGrid electrodes.

## 3.2 Component Integration

To integrate the components, three key factors had to be considered. First, the key components that enable the integration (the connector) had to be assembled. The resources (materials and instructions) for these steps were publicly available through the website of the cEEGrid developers and only required the order of two very affordable components (see Appendix A.2 for more detail). The overall cost for the purchase of these system components was below 30 USD. Second, a mounting solution had to be devised that secures the connector cables (to reduce recording artefacts from cable movements) and the PCB board together with the OpenBCI boards (to create a single, cohesive component). This step represented the major challenge in the integration process as several requirements had to be met – primarily the requirement of keeping the effort and required resources for the assembly to a minimum. For this reason, all designed mounting parts only require materials that are already shipped with the OpenBCI or require regular fused deposition modeling (FDM) 3D printing (the same way that OpenBCI components can be printed already). The mounting solution was designated to be versatile; that is, it was conceived to be used with different placements on the body or different headgear. Therefore, in a first step, a holding clip was designed that can be screwed on to the bottom of the already available Cyton board holder (see Figure 3). Through this clip solution, the OpenBCI boards are mountable on headbands, baseball caps, or similar headgear. It was also considered that, depending on the length of the (interchangeable) cables, the board could also be placed on scarves, collars, shirts, or belts. However, it was found that the weight of the cEEGrid PCB required structural support closer to the ears, which is why such a body-based application is not easily achievable with the current design. Similar challenges to such ear-based EEG recordings have previously been solved with more stable holders that support the weight of the system components directly around the cables (see, e.g. [45]). Another requirement to the mounting solution was that the cEEGrid PCBs should be easily attachable to the OpenBCI boards to form a single cohesive recording unit that does not interfere with regular use of the OpenBCI headset. Therefore, a PCB holder was designed, that can be 3D printed together with an altered version of the Cyton board cover and can be easily clipped between the Cyton and Daisy boards (see Figure 3). Thus, if a researcher wants to switch from a cEEGrid to a regular OpenBCI EEG recording (e.g., using the Ultracortex Mark IV headset), the PCB holders can easily be removed. Also, if a researcher wants to use the Ultracortex headset together with the cEEGrids, this design supports it.

Third, in a last and critical step, a solution had to be devised to flexibly connect the cEEGrid PCB to the Cyton and Daisy boards. The flexibility was required primarily because the OpenBCI board can only connect to 18 out of the 20 cEEGrid electrode channels (including reference and ground electrodes). Initially, jumper cables were directly soldered to the cEEGrid PCB. However, this solution was discarded because, in this form, one electrode cable was always free and would have to be additionally stored. With this solution, it was also difficult to correct errors in wiring later on. Therefore, pin headers were soldered on to the PCB (see Figure 3). This approach is not only in line with the OpenBCI design philosophy, but it also enables the flexible use of different cable lengths and electrode mappings, and it enables easy visual control of the correctness of mappings. As one electrode per ear needs to be left out in the recording, the mapping of electrodes to the Cyton and Daisy board can become a bit challenging at first. For this reason, we have developed a simple schematic (see Figure 4) that visualizes how PCB channels and Cyton and Daisy pins can be connected while maintaining a consistent color-coding of channel numbers for both ears.

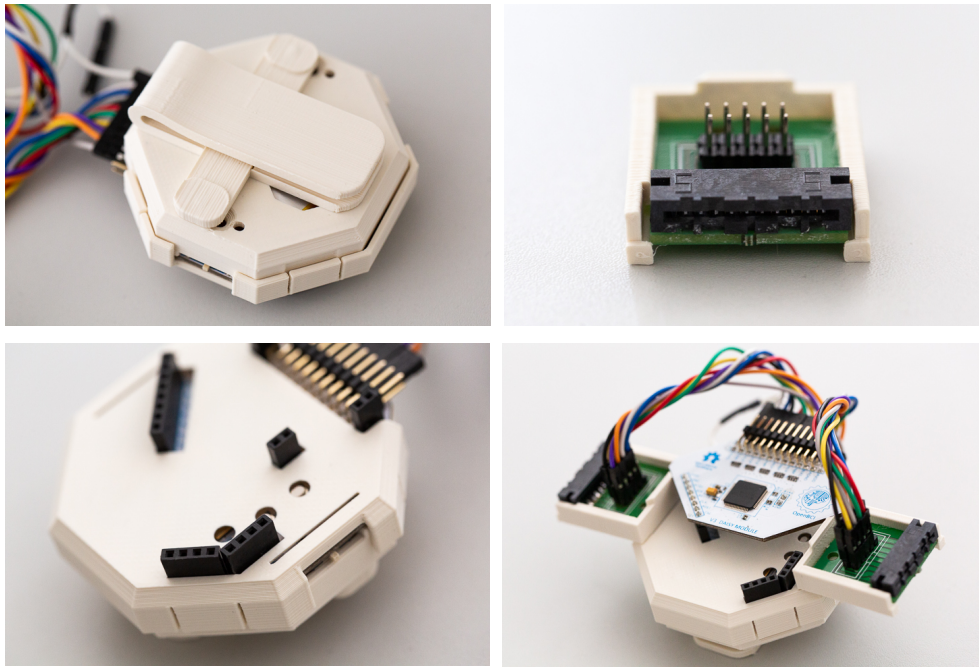

Figure 3: The (re-)designed components for the system integration. Top left: The new 3D-printed clip that can be screwed on to the bottom of the regular Cyton board holder. Top right: The 3D-printed PCB holder. Bottom left: The altered OpenBCI Cyton enclosure to which the PCB holder can be clipped on to. Bottom Right: The assembled components as a single unit.

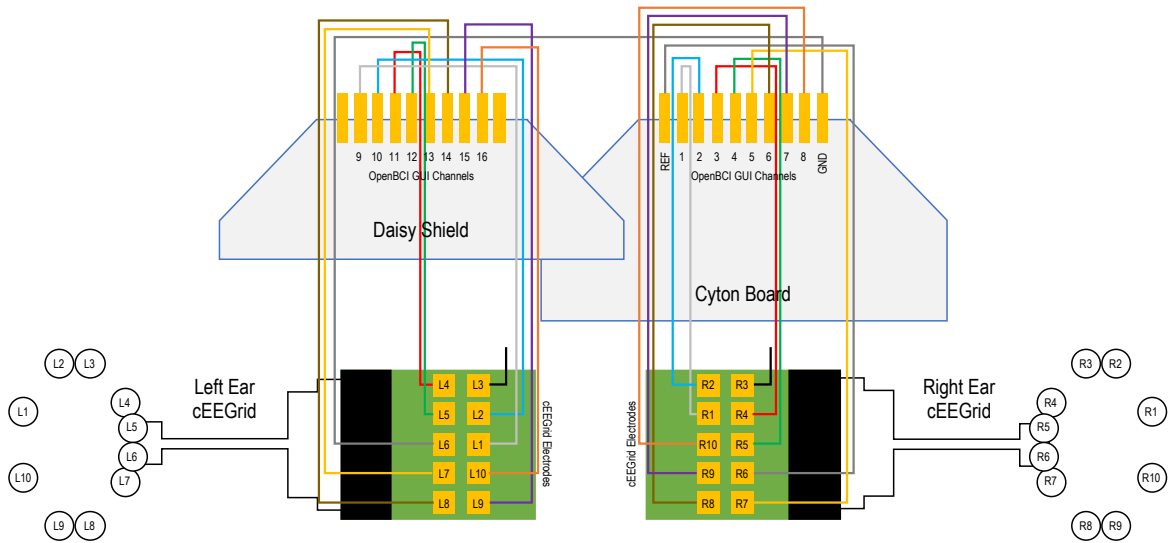

Figure 4: Schematic of the mapping of cEEGrid electrode channels on the PCB to the Cyton and Daisy pins, including a color-coding and channel reference as visible in the OpenBCI GUI during recording.

Given these assembly possibilities, a researcher needs to decide which electrodes to include or exclude and which to use as reference and ground electrodes. Previous research has used the R5 and R6 electrodes as reference and ground electrodes [22,53,60,69]. For the sake of keeping the color coding consistent, we have here opted for the use of L6 and R6 as ground and reference, respectively. This configuration allows to similarly compute a linked mastoid reference (e.g., using L5 and R5) as in previous cEEGrid work. In the herein outlined configuration, we have chosen to leave out electrodes L3 and R3, as we were interested as to whether or not other electrical signals from the body can be collected that might be useful in feature extraction and artefact removal processes later on. However, if, for example, a researcher is more interested in collecting more data from sources closer to the brain, they might want to decide to exclude electrodes below the ear instead (e.g., L8 and R8). We have explored recordings with both configurations and found that the system effectively captures typical EEG phenomena, with only minor differences. To demonstrate the recording feasibility with the described configuration, the following chapter details the results from a feasibility study.

## 4 STUDY DESIGN

### 4.1 Goals & Materials

To ascertain the potential of using the cEEGrid-OpenBCI integration for HCI research, the goals of this experiment were threefold, namely to collect data that would: (1) indicate the presence of well-known and easily detectable neural processes, (2) serve the replication and extension of relatively young observations of cEEGrid applicability, and (3) allow the exploration of new cEEGrid application potentials in the HCI domain.

For the first goal, as an essential reference, visual stimulation was manipulated in the form of eyes-open and eyes-closed resting phases. This was done to observe the detectability of the so-called "Berger effect" (posterior Alpha frequency power increases when the eyes are closed, and no more visual sensory input is processed in the occipital brain regions – see, e.g. [11,22]). The Berger effect has already been confirmed to be well-visible in the cEEGrid electrodes by the electrodes' developers [22]. In our experiment, participants were asked to either keep their eyes closed or to keep them fixed on a cross in the middle of the screen for one minute. To further assess if intensified visual stimulation could be detected through the cEEGrids, participants were asked in a second eyes-open phase to watch a video of fish swimming in the ocean. This stimulus has in previous research been used to conduct eyes open resting phases [63].

For the second goal, task difficulties were manipulated to explore the potential of the cEEGrids to differentiate situations of low, moderate, and high mental workload. For mental workload, there has so far only been one report that such differentiations are possible in driving simulations with older adults [69]. However, this study also reported sensitivity complications possibly related to broadband artefacts. In the present experiment, a mental arithmetic task was chosen in which participants had to sum up numbers with different amounts of summands to alter the task difficulty (easy, moderate, and hard difficulties – for a reference to the task, see [39]). Also, a time production task was included in the experiment to provide an alternative task that requires low mental workload (and therefore, to assess the sensitivity of cEEGrid recordings for workload changes). In the time production task, participants were asked to press a button after a few-second period had passed following a notification sound (see, e.g., [6]). It has been documented that participants typically tend to count up the time in their heads, which leads to focused concentration, yet low mental load [6]. To collect a ground-truth measure for mental workload, participants responded to the NASA-TLX questionnaire [33] that captures workload through six dimensions (mental demand, physical demand, temporal demand, performance, effort, frustration) on a rating scale of 0 (lowest) to 21 (highest) after each experiment condition. Furthermore, for the second goal, both to assess the recording system's feasibility and extend the knowledge on which features can be extracted from the cEEGrid recording, an ECG was recorded in Lead II configuration (using a Biosignalsplux amplifier with 1000Hz sampling frequency and 16-bit resolution). This measurement was included because previous work has documented the possibility that electrical heart activity is collected by the cEEGrids [14,60]. Yet, no assessment of the completeness of the ECG trace has been conducted.

For the third goal, relationships between recorded EEG data and reports of the so-called flow experience were assessed. The flow experience is defined as the unique, holistic sensation people experience when they act with total involvement [21]. In this state, people experience complete absorption into a task and highly fluent action, a situation in which everything appears to be under perfect control [25]. The flow experience is of heightened interest for developers of adaptive games (e.g. [27]), for designers of digital social interaction experiences (e.g. [37,44]), and for digital professionals (e.g. knowledge workers or eSports athletes – see [64]) because it has often been linked to superior performances, personal and social growth and heightened general well-being [68]. It is not yet possible to continuously and unobtrusively observe flow experiences (i.e., without interruption from surveys) [42]. However, such an ability would open up the possibility to learn about flow experience dynamics and interventions that foster the experience. Besides validating the OpenBCI-cEEGrid system integration, the present study explored the potential of flow detection through cEEGrids. This approach was considered promising, as some previous research on flow neurophysiology has suggested a possible relationship between flow experiences and verbal-analytic reasoning processes. Specifically, it has been considered that athletes (e.g. shooters, archers, and golfers) might experience more intense flow due to lower interference of verbal-analytic reasoning during the execution of a learned motor behavior sequence (less inner verbal correction during movement) [30,70]. As temporal brain regions (whose activity is primarily collected by the cEEGrids – see [47]) contain such verbal-analytic processing regions [30], the observation of increased/decreased activity in these regions by observation of Alpha power changes could be a way of detecting flow experience intensities. To enable the exploration of this pattern in our experiment, the mental arithmetic task has been designed not only to elicit different mental workloads but in alignment with previous work, to provide the entry conditions for flow through performance-adaptive difficulty in the moderate difficulty condition (this design was adapted from previous flow research – see [39,42]). Performance-adaptive means that the difficulty increased one level after two successful trials in a row and decreased one level after an incorrect trial, thus continually balancing the task difficulty. To collect a ground-truth measure for flow experience, participants responded to short flow surveys after each mental arithmetic task condition. Three items from the Flow Short Scale (FKS) by [25] were used with one item each for the factors fluency, absorption, and optimal challenge each (item wordings are shown in Figure 11). Participants reported their (dis-)agreement with all items on a scale from "Not at all" (= 1) to "Very much" (= 7).

## 4.2 Study Execution

In summary, the experiment was comprised of situations of rest (eyes open and closed), of low task demand (time production and mental arithmetic), of moderate and adaptive task demand (mental arithmetic), and of high task demand (mental arithmetic). In line with general EEG research recommendations, each stimulus condition was presented repeatedly, specifically twice in this experiment. The order of the conditions was fixed. The entire procedure is visualized in Figure 5.

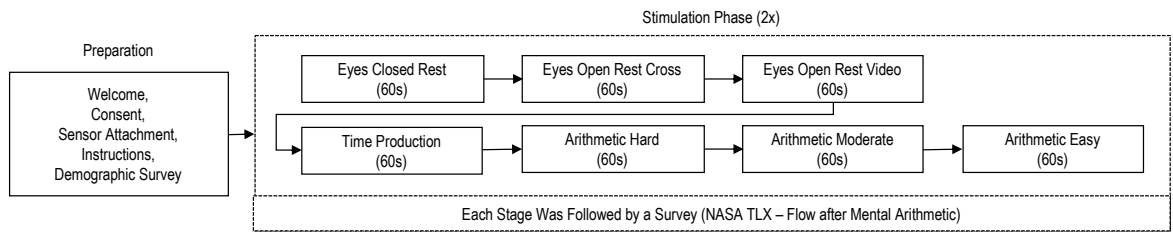

Figure 5: The experiment procedure visualized.

Data were collected for six male participants in the age range of 24 to 30 (mean = 27.83, median = 29.5). All participants were right-handed and had full eye- and color-sight (with or without correction). Participants were also screened for being generally healthy and not taking any mind-altering medication. Participants were sampled from colleagues at the academic institution of the first author and participated without remuneration. The experiment followed the ethical guidelines from the first author's academic institution, and participants provided informed consent before their participation. The experiment was conducted in an office room belonging to the institute of the first author. Upon arrival in the recording room, participants were first informed about the type of recording and signed the consent form. Afterwards, gelled ECG chest electrodes and the gelled cEEGrids were attached to the participant. For the cEEGrids, alcohol was used to clean the area surrounding the ear. Next, the EEG signal quality was assessed using the OpenBCI GUI software. Specifically, good electrode contact was assessed by looking for typical amplitudes (ca. 5-10 µV) and the absence of high-frequency power abnormalities in the OpenBCI GUI FFT widget. Afterwards, the experimenter left the room, and the experiment was completed fully autonomously by the participants.

## 4.3 Data Pre-Processing

EEG data were processed primarily following the previous cEEGrid work of [22,53,60,69]. A fully automated data preparation process (with descriptive statistical and visual inspection of the data before and after processing) was implemented to make the feature extraction process transparent and reproducible (see, e.g. [8]). The full processing pipeline, including the parameters for each step, is documented in Appendix A.1. ECG data were processed according to the Pan-Tompkins method to reduce the baseline wander, filter out muscle artifacts and other high-frequency noise components, and perform R peaks detection [61]. Then, the R-R periods (i.e., the time distance between consecutive heartbeats) were computed. For one participant, all of the ECG data had to be discarded as the electrodes came off during the recording. Questionnaire variables were aggregated through their recommended approaches, that is, the six NASA-TLX items were summed [33], and the three flow items were mean averaged [25].

# 5 RESULTS

## 5.1 Changes of EEG Frequency Band Powers with Conditions

To provide an initial overview, distributions of the EEG frequencies are reported in the form of power spectral density (PSD) distributions for each condition. These allow assessing changes in frequency bands by experimental condition visually. Figure 6 shows the grand average PSD traces for all experiment conditions (average of all electrodes and participants).

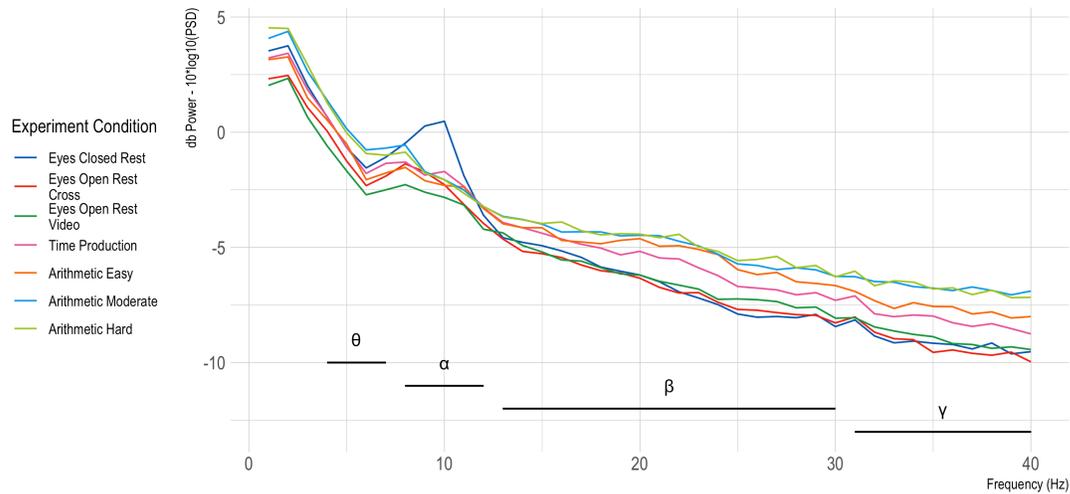

Figure 6: Grand average PSD distributions showing changes in EEG frequency power by experiment condition. Extracted frequency band ranges are shown by black lines below the PSD.

To further assess condition differences, statistical analyses were conducted in the form of linear mixed models (LMM) with random intercepts that account for inter-individual variances. Post-hoc contrasts with Bonferroni correction were used further to identify pairwise condition differences taking inflation of false-positive errors from multiple testing into account. Figure 7 shows the ANOVA test results combined with the post-hoc contrasts in a pairwise p-value plot for all four extracted frequency bands. Significant differences across the conditions are found for all frequency bands with stronger contrast in the lower frequency ranges (Theta and Alpha). These differences are described in the following sections.

## 5.2 Detection of Classical Neural Patterns – The Berger Effect

The presence of the well-known Berger effect (increased Alpha power during eyes closing) is visible in the peak in the 8-12 Hz region (see Figure 6), a result that is in line with previous cEEGrid work [22]. An ANOVA based on an LMM with Alpha power (8-12 Hz) as the dependent variable, condition as the independent variable, and participants as random intercept further shows that a significant main effect is present (see Figure 7). The Bonferroni-corrected post-hoc tests further indicate that the Alpha power changes in the eyes closed phase are significantly different from all other phases (see Figure 7). This finding demonstrates the expected recording feasibility of the presented system. It was considered that higher visual stimulation from watching a video compared to a static image of a fixation cross might result in further reduction of Alpha power. This effect is indicated in the PSD, but not found as a significant difference. This result could mean that the effect size might be too small or that further feature extraction finesse is required. Nevertheless, the clear replication of the Berger effect provides evidence for the technical feasibility of the OpenBCI-cEEGrid integration.

## 5.3   Extension of Young cEEGrid Observations – Mental Workload

Additional evidence for the system feasibility comes from the observed changes in Theta (4-7 Hz), Beta (13-30 Hz), and Gamma (31-40 Hz) power. The PSD distributions in Figure 6 indicate that both low- and high-frequency powers show increases with more difficult tasks, a known finding from neuroscientific literature (see, e.g. [50]). For cEEGrid research, increases of Theta power with higher mental workload have also been reported in a first and recent study [68]. For all three frequency bands, the ANOVA test based on an LMM shows a significant main effect (see Figure 7). The post-hoc contrasts show that significant ($p<0.001$) Theta power changes are visible between higher mental workload tasks (i.e., Arithmetic Hard and Arithmetic Moderate) and rest phases and, to a weaker degree, between lower mental workload tasks (i.e., Arithmetic Easy and Time Production) and rest phases. Regarding the latter, the conservative Bonferroni correction should be considered.

To extend these analyses, relationships between frequency band powers and self-reported mental workload were assessed. For this purpose, NASA-TLX reports were aggregated (responses for all six dimensions summed) to form a composite indicator of perceived mental workload. Figure 8 shows the distributions for this variable and the results from an LMM with post-hoc contrasts. The results indicate that mental workload was lowest in the resting phases (eyes open and closed), increased slightly with the time production and easy mental arithmetic task (no difference between the two), and increased further stepwise for the moderate and hard mental arithmetic tasks. Afterwards, to take the individual baselines for reported workload and physiological response into account, all data were subsequently z-standardized first and submitted to linear regression models with the reported workload as the dependent and average Theta, Alpha, Beta, or Gamma power as independent variables. This approach provides a multilevel analysis similar to the previous tests here and related HCI-neurophysiology research (e.g. [32,46]). The eyes-closed condition was excluded for this analysis to account for the influences of eyes closing on Alpha power changes (the Berger effect). Figure 9 visualizes each linear model. Significant, positive relationships are found for Theta, Beta, and Gamma band power, but not Alpha band power. The results from this analysis further indicate a possibility to infer levels of experienced mental workload to a moderate degree (given the $R^2$ coefficients ranging from 0.108 to 0.278) and further suggest that typical observations of reduced Alpha power with higher mental workload (see, e.g., [17]) are not as visible, possible due to confounding influences over these temporal regions. This effect might be masked by diverging effects within the Alpha frequency band (see, e.g., [41,50]).

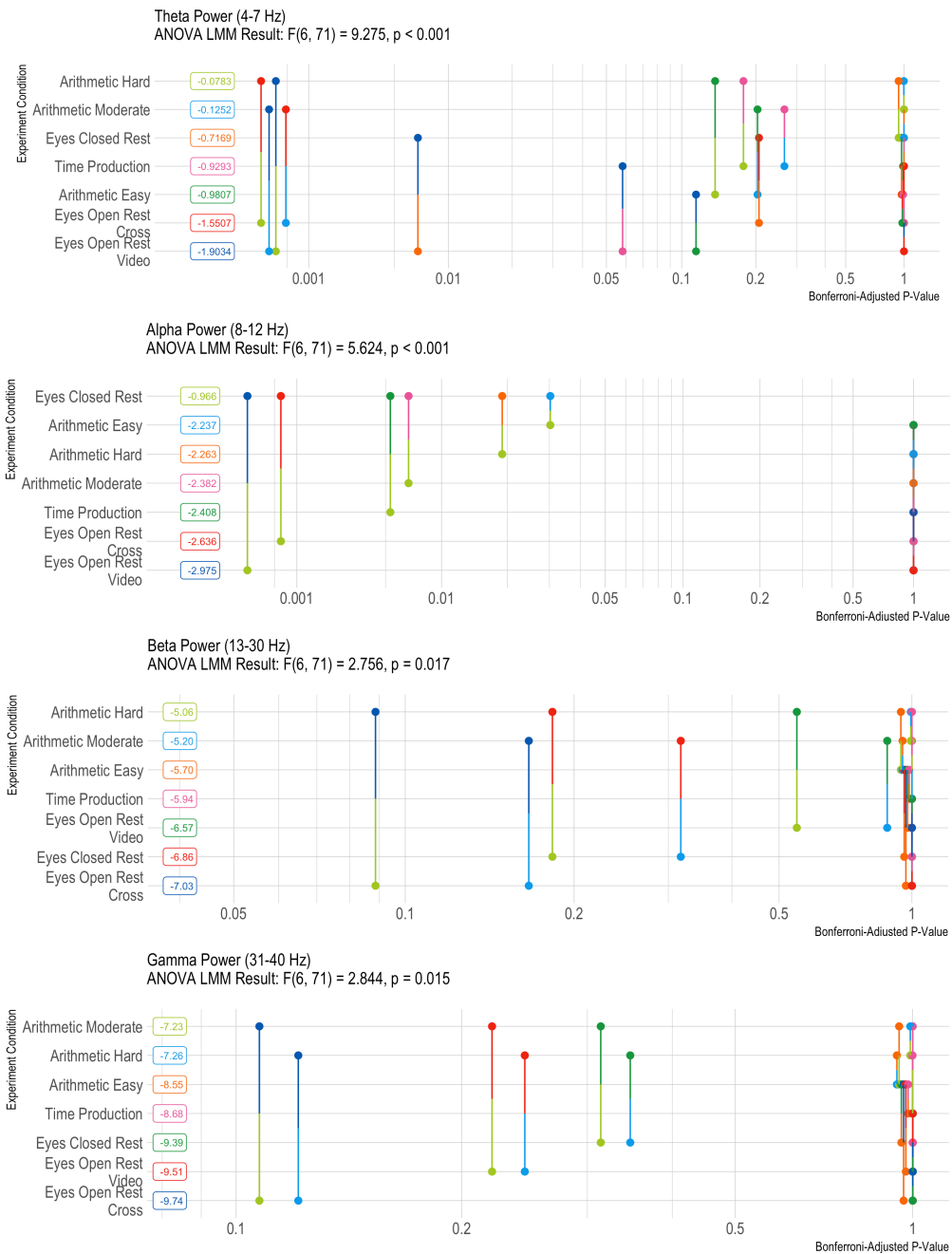

Figure 7: Visualization of the post-hoc condition comparisons by a pairwise p-value plot for Theta, Alpha, Beta, and Gamma frequency bands, including the test statistic from an ANOVA on an LMM with each band as dependent variable.

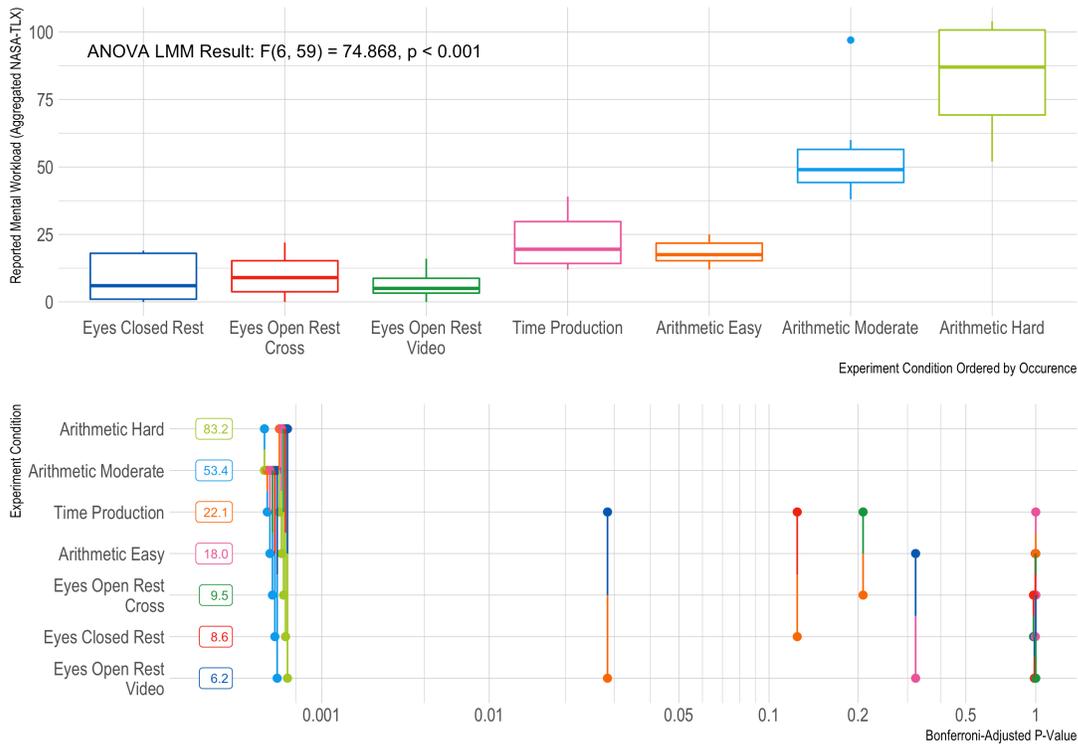

Figure 8: Changes in reported mental workload by experiment condition. Top: Boxplot distributions including the test statistic from an ANOVA on an LMM with reported workload as dependent variable. Bottom: Visualization of the post-hoc condition comparisons by a pairwise p-value plot.

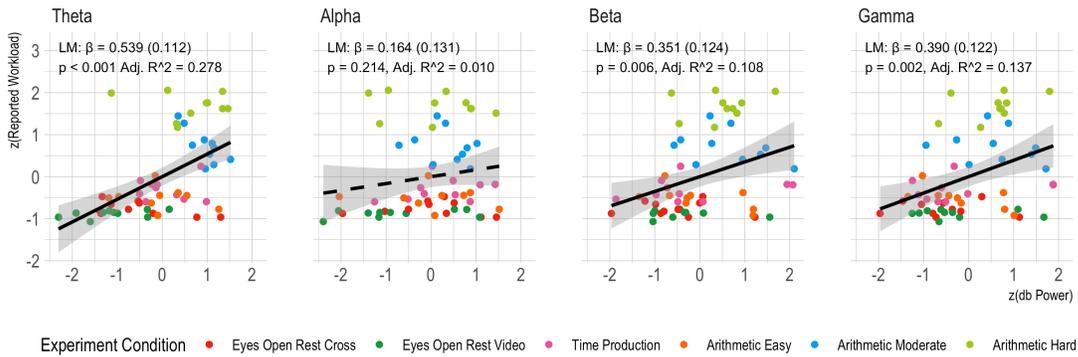

Figure 9: Relationships of reported mental workload and frequency band power (both z-standardized) analyzed in a linear regression model. The grey ribbon represents 1 SE. A dashed line represents an insignificant regression coefficient.

Altogether these findings suggest a moderate degree of sensitivity for detecting different levels of mental workload that is potentially weaker than, for example, the differentiation of workload levels over frontal or whole scalp regions (see, e.g., [5,27,50]). This finding is in line with the previous results by [68] that showed some, but limited, possibilities of differentiating workload levels through cEEGrids. However, a novel observation could be the difference in Theta power between resting phases and tasks with low mental workload. Though not statistically significant ($0.05<p<0.15$), such difference might indicate the possibility to distinguish situations of no task concentration (possibly mind wandering) from tasks that require even mild task-directed concentration. This represents a novel observation for cEEGrid research that should be investigated further.

### 5.4 Extension of Young cEEGrid Observations – ECG Extraction

Given that previous research has indicated that the cEEGrid electrodes might be collecting more physiological information than just changes in neural activity, specifically, the electrical activity of the heart [11,60], it was explored how accurately this feature is represented by comparing the EEG recording to a standard ECG recording in Lead II configuration. For this EEG-ECG signal comparison, ECG traces extracted from independent components (IC) of the cEEGrid recordings were first visually compared for their similarity to regular ECG recordings. Figure 10 shows an example of the selected (IC) for each of the remaining participants together with a regular ECG signal from the same resting phase (to facilitate morphology comparisons). From these traces, the presence of the characteristic R wave is visible, particularly in four out of five participants, which further confirms the feasibility of the herein presented recording system. Interestingly, it has been reported that only for 50 to 60% of participants, this ECG signal trace can be retrieved [11]. Our sample is very small, yet it might indicate that the percentage in the population for which this ECG trace is available in the cEEGrid data could be a bit higher. Nevertheless, from top to bottom, the traces show a reduction of signal-to-noise ratio and increasingly less evident R-peaks, reflecting variability in subjects' physiological characteristics, electrodes adherence to the skin, or small changes in their placement.

To further assess the quality of these cEEGrid-based ECG derivations, the R-R sequences extracted from both signals were compared using the Bland-Altman approach. This method assesses the difference between a reference measure (i.e., the ECG-derived R-R periods) and an alternative one (the IC-derived R-R periods) through statistical and graphical tools [9]. The Bland-Altman statistics suggest that the cEEGrid-derived R-R periods were highly comparable (mean absolute difference: 1.6 ms; 95% Gaussian Limits of Agreement (1.96*SD): ± 81.0 ms; 95% non-parametric Limits of Agreement (1.96*IQR): ± 9.8 ms) and significantly correlated (Pearson correlation coefficient: 0.94) with the reference ECG R-R periods. In particular, the notable difference between Gaussian and non-parametric LOA indicates that the cEEGrid-derived R-R periods correctly match most of the reference ones, and the few that are not matched are likely due to rare erroneous R waves detections on the selected IC. Since such erroneous R-R periods may be easily excluded by performing basic outlier detection, this observation further supports the reliability of the cEEGrid-derived R-R periods as a good measure of cardiac variability and the potentiality of reliably retrieving a useful ECG trace from the cEEGrid recordings, especially for non-clinical applications where slight imprecision is acceptable.

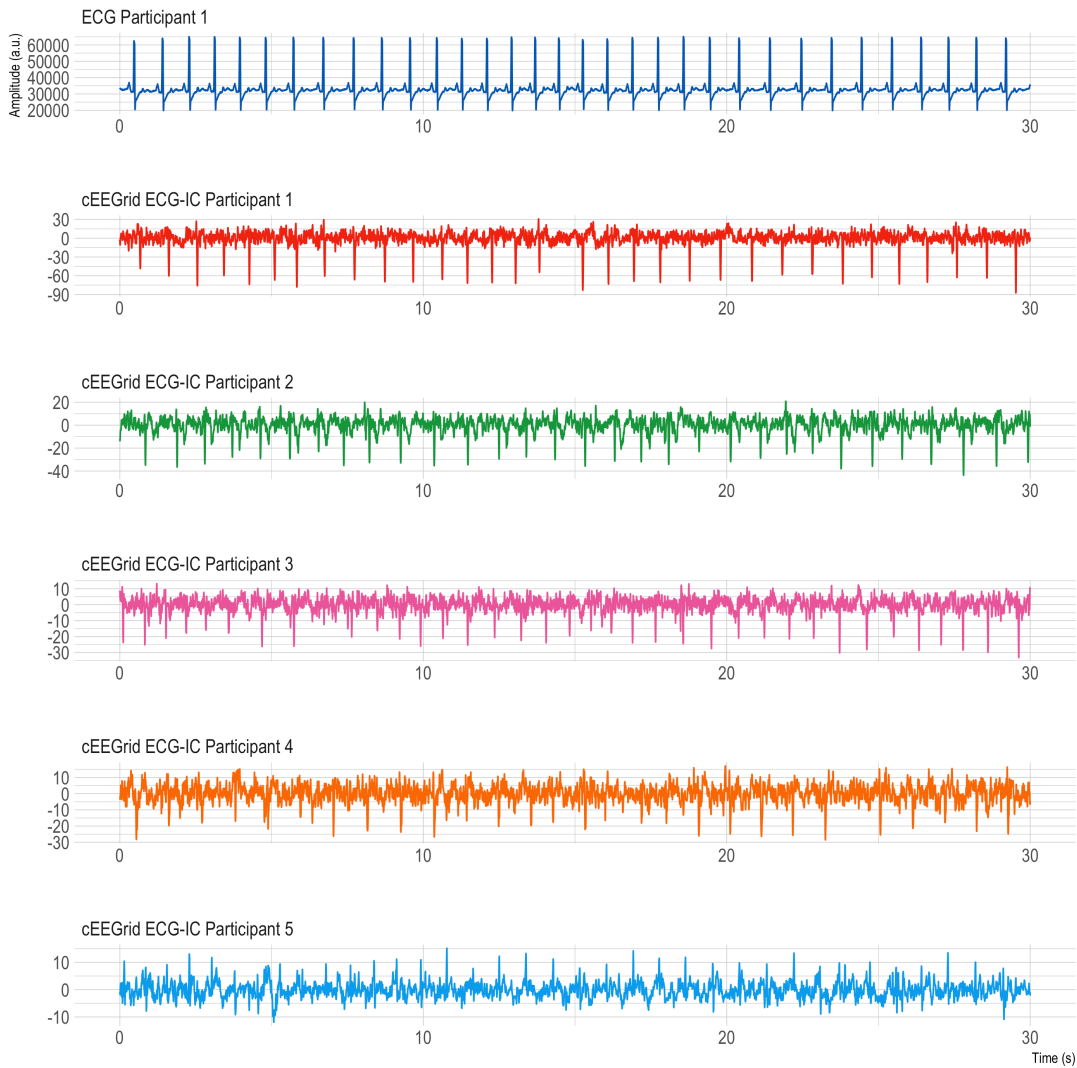

Figure 10: Traces of a Regular ECG (Top Row) Together with cEEGrid-derived ICs Selected For Each Subject Due to Their Visual Similarity to a Regular ECG Trace. The Shown Signals Belong to the Eyes Closed Rest Condition.

### 5.5 Exploration of New cEEGrid HCI Application Potentials – Flow Experience Detection

To explore the HCI-application possibilities of the cEEGrid-OpenBCI extension, relationships between flow experiences and EEG frequency band powers were investigated. The flow experience report distributions (mean score and individual items for optimal demand, fluency, and absorption) for all three mental arithmetic conditions are shown in Figure 11, together with the results from an ANOVA test based on an LMM. As intended by the task design, flow experiences were heightened during the moderate and adaptive workload conditions, for which optimally balanced difficulty was reported. In line with related flow research, reduced fluency is reported during the hard difficulty, and reduced absorption is reported during the easy difficulty condition (see, e.g., [61,66]). Thereby, the elicitation of experience variation was deemed supported, and the evaluation of flow and temporal EEG power relationships pursued.

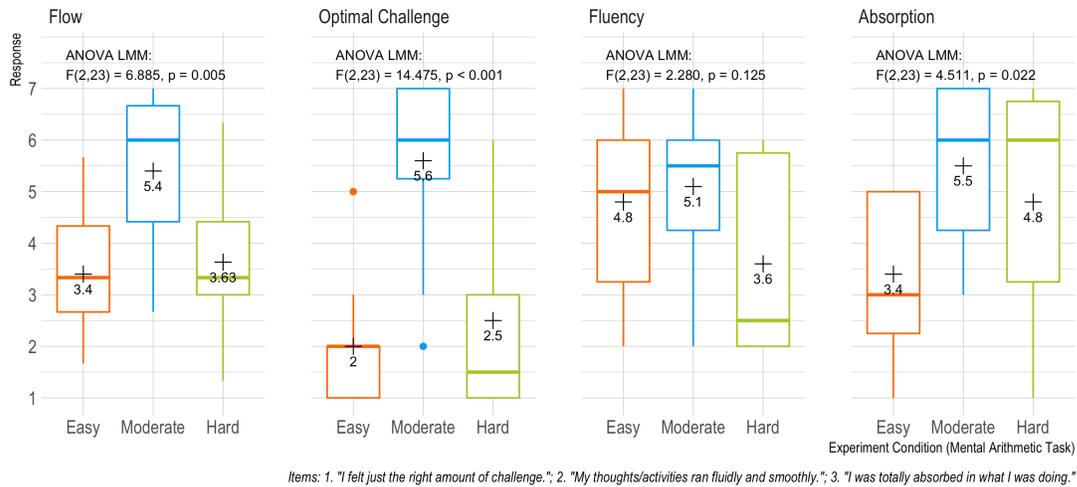

Figure 11: Distributions of reported flow experiences for the mental arithmetic task (mean of three items and individual items for optimal challenge, fluency, and absorption) including the test statistic from an ANOVA on an LMM with self-reports as dependent variable. Crosses and numbers next to them represent means.

As with the workload analyses, all data were subsequently z-standardized to take the individual baselines for reported flow and physiological response into account. The data were then submitted to linear and quadratic models with the reported flow as the dependent and average Theta, Alpha, Beta, or Gamma power as independent variables. Quadratic relationships were explicitly used here for the reason that recent flow research has increasingly found non-linear patterns between flow reports and physiological variables [7,66]. The model results are shown in Figure 12. Theta, Beta, and Gamma activity showed no significant relationship with flow. Alpha activity, on the other hand, showed a significant quadratic relationship with the reported flow intensity. To further examine this relationship, similar quadratic regression models were also created for the individual flow items. Figure 13 shows these results and highlights that fluency perceptions do not appear to relate to this Alpha power change, but that optimal challenge perceptions and absorption levels do. These results suggest that more intense flow experiences (particularly experiences of challenge and absorption) are likely accompanied by moderate levels of temporal Alpha activation. This represents a novel observation for cEEGrid and flow-related HCI research that should be investigated further. Possible interpretations and avenues for future research are discussed next.

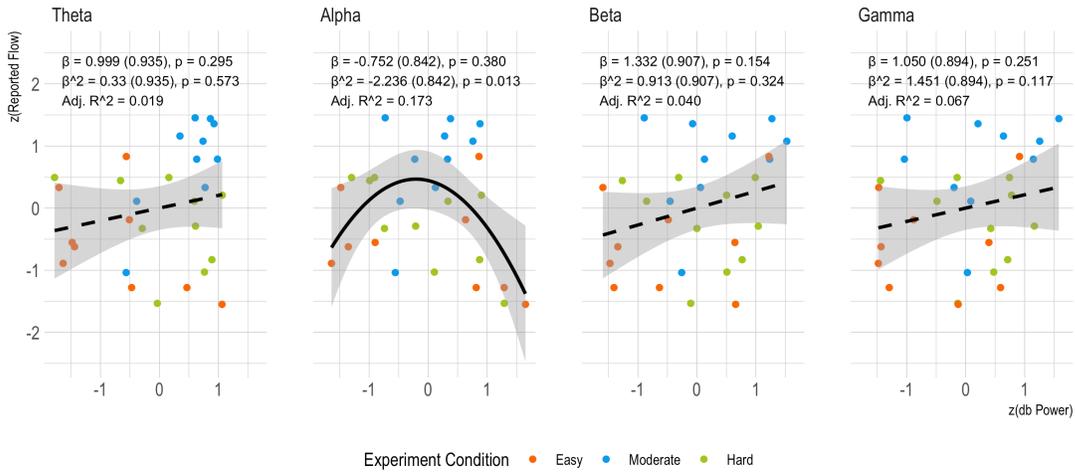

Figure 12: Relationships of reported flow and frequency band power (both z-standardized) analyzed in a linear or quadratic regression model. Quadratic models used orthogonal contrasts. The grey ribbon represents 1 SE. Dashed Lines indicate an insignificant regression coefficient. For Theta, Beta and Gamma, linear model fits are shown due to indication of better model fit.

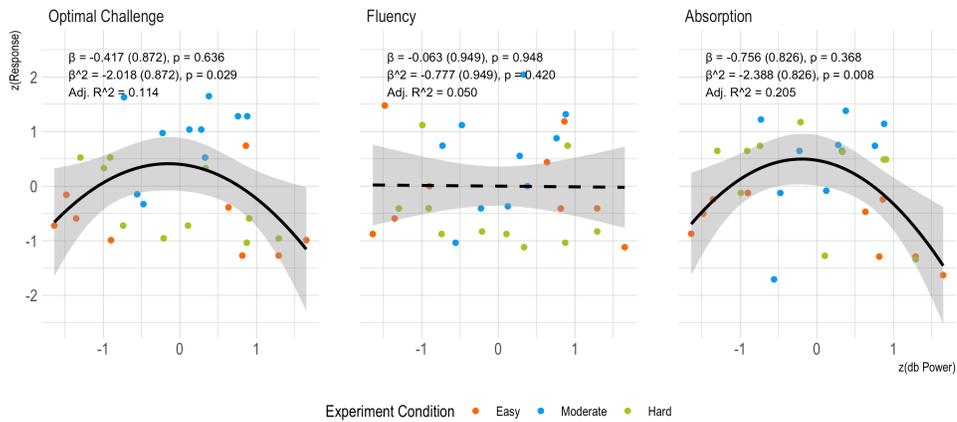

Figure 13: Relationships of reported challenge, fluency and absorption with Alpha band power (both z-standardized) analyzed in a linear or quadratic regression model. Quadratic models used orthogonal contrasts. The grey ribbon represents 1 SE. Dashed Lines indicate an insignificant regression coefficient. For fluency, a linear model fit is shown due to indication of better model fit.

## 6 DISCUSSION & CONCLUSION

In this work, we demonstrate the feasibility of extending the OpenBCI biosignal acquisition platform to the use of a novel around-the-ear EEG sensor array (cEEGrids). Despite the general possibility of using cEEGrids with various amplifiers [22], this technical validation is an essential contribution as the cEEGrids present a challenging setup for EEG amplifiers [47]. The results from a controlled experiment show that the system records well-known and expected neural changes from visual stimulation and mental workload, and also the possibility of extracting ECG signals from the EEG recordings. Doing so, the analyses for the latter two aspects further support the suitability of cEEGrids for measuring these variables, yet with some caveats. As in a previous study [68], observations of changes in mental workload appear to be only reflected in EEG frequency powers with moderate sensitivity. Important for HCI applications is the indication that even low increases of workload might be separated from rest. Therefore, at least three classes of workload experiences might be separable (one of which likely represents a lack of task-directed concentration), which could be useful for the development of productive adaptive systems for knowledge workers (e.g., software engineers, designers, programmers). Additional valuable information for such systems could also be derived from heart rate variability (HRV) metrics that could be used to infer levels of physical activation that are known to be conducive to performance and health (see, e.g., [40]). Comparing the EEG-ECG components to a regular ECG recording further confirms that the cEEGrid recordings collect valuable information on heart rate changes. While this has been reported previously [11,60], no quantification of the accuracy of this observation has been conducted, and we, therefore, presented a novel analysis that suggests a possibly increased potential for ECG extraction from cEEGrids. The (sparse) data from this experiment not only demonstrate that an ECG trace with high similarity to a regular ECG recording can be extracted from the cEEGrid recordings. With four out of five participants showing a good extractability, further study should be conducted to learn for how many and which individuals this interesting additional information can be collected in the process of cEEGrid recordings, without the hassle of using a separate device to acquire it. Further study is required not only because the participant pool was small in this instance, but also because some of the outlined effects suffer from a lack of specificity. This might be overcome by applying more dedicated and sophisticated signal processing techniques (see, e.g., [12,16]) or by leveraging more sophisticated feature extraction methods that select critical electrodes, frequencies, and timings (see, e.g., [57,68]).

In the last analysis, the HCI-application potential for cEEGrids was further explored by studying the relationships between flow experiences and temporal EEG frequency band power. This was considered a possibility as previous flow neurophysiology research has suggested a possible relationship between flow experiences and verbal-analytic reasoning processes (that could be observed through temporal Alpha power changes – see [30,69]). The results in this experiment do suggest the presence of a link between temporal Alpha power and reported flow experiences (and, in particular, the level of task absorption – a flow subdimension – see [26]). As high levels of Alpha power show a reduction of neural activity (Alpha is an inhibitory frequency – see [41]), the herein observed pattern might best be explained as an efficient use of verbal-analytic reasoning activity during the mental arithmetic task. Thereby, a low level of Alpha power is interpreted as a state of "over-thinking" the task because demands might be too high so that cognitive strategies to cope with the difficult task are being engaged. A high level of Alpha power, on the other hand, might suggest a lack of verbal-analytic engagement in this task, a state of "under-thinking," as the task demands are, for example, too easy or disengagement from the task has occurred. This interpretation would be coherent with the observation that particularly absorption experiences are linked to this inverted U-shaped Alpha power pattern. Specifically, both the low and high Alpha power situations would appear likely to show low absorption then, as one would not be absorbed in the task when being disengaged ("under-thinking") or being overly concerned with the development of task strategies ("over-thinking") and thus focused on internal processes and not on the task itself. Together with the findings for workload and ECG extraction, these results represent an interesting avenue for the HCI domain as they provide new means to observe the emergence of flow experiences through possible combinations of Theta power observation (workload & concentration), HRV observation (physiological activation), and Alpha power observation (verbal-analytic reasoning). The flow experience is of heightened interest for developers of adaptive games (e.g., [27]), for designers of digital social interaction experiences (e.g., [37,44]), and for digital professionals (e.g., knowledge workers – see [63]) due to the experiences links to individual and social performance, growth and well-being [67]. Again, as the sample in this work is very small, this relationship needs further investigation. Initially, such future work could combine the established difficulty manipulation paradigm across various tasks (mental arithmetic and simple games like Tetris or racing games before – e.g., [27,66]). Beyond these validations, it would be of particular interest to observe natural situations of flow experiences. A promising avenue for this work would be to study the experiences of eSports athletes. This subject group combines the advantages of a motion-reduced workplace (to circumvent currently active issues of dealing with motion artefacts in cEEGrid recordings – see [60]) with high levels of task expertise. These two factors that are likely to be conducive to EEG-based flow research.

A final compelling direction that the herein presented research enables is to concurrently use the cEEGrids with the OpenBCI Ultracortex headset (see Figure 14). So far, no other systems enable the simultaneous recording of cEEGrid data with other electrodes in a single system. By leveraging the herein presented design, a researcher could easily collect data on some cEEGrid electrodes and some electrodes on the scalp to explore which observations can be made from more distributed recording sites and alternative reference montages. Thereby, it is, for example, greatly facilitated to identify essential electrode positions for the observation of a particular neural activity pattern and to develop prototypes for wearable recording systems. Altogether, we, therefore, hope that this work contributes to further research on BCI and HCI research and demonstrates the exciting possibilities that this new system integration offers for this purpose. It is our firm belief that both access to research-grade neural activity recordings and increase recording acceptability (due to higher comfort and less obtrusion) will be a key driver on the way to enabling daily-use BCI applications.

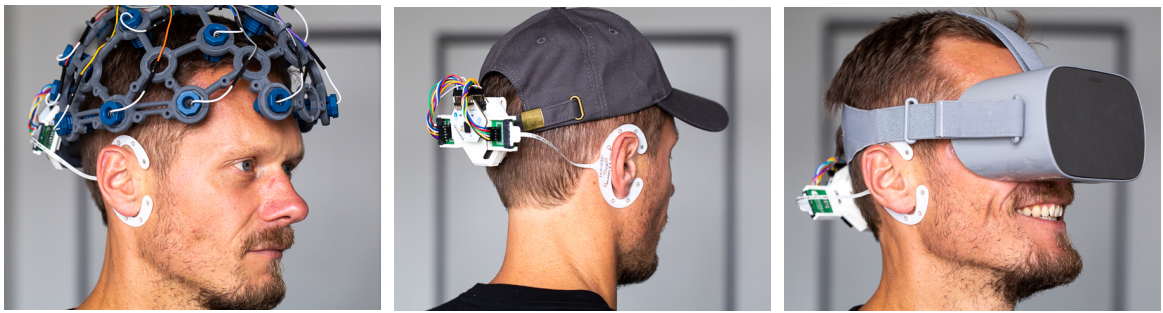

Figure 14: Different mounting and application scenarios for the OpenBCI board with cEEGrids. Left: OpenBCI Ultracortex Mark IV together with cEEGrids for investigations with combined sensors. Middle: cEEGrids Attached to a baseball cap for wearing in field study situations. Right: cEEGrid observations in VR experiments through direct attachment of the OpenBCI board to the VR head strap.

# A APPENDICES

## A.1 EEG Signal Processing Pipeline

The EEG signal processing was conducted using R & RStudio (initial inspection and cutting of the data), and the Matlab toolbox EEGLab [24] Version 2019.1 together with custom Matlab scripts.

Table 1: Steps and Parameters of the automated EEG Signal Processing Pipeline

| # | Step | Parameters | Reference |
|---|---|---|---|
| *Inspection & cutting (R)* | | | |
| 1 | Cutting data into segments for each experiment condition | All data during condition extracted | – |
| 2 | Descriptive statistics for cutting integrity assessment | • Expected and actual nr. of extracted samples<br>• Correctness of first and last timestamp per phase | – |
| *Data cleaning & main feature extraction (Matlab)* | | | |
| 3 | Baseline correction | Channel mean subtraction | [53] |
| 4 | Re-referencing | Algebraically linked mastoids | [22] |
| 5 | Remove line noise | CleanLine EEGLab plugin<br>• 25 & 50 Hz components | – |
| 6 | Detrend | Windowed FIR<br>• 1Hz high pass filter<br>• Hann windows with 50% overlap<br>• Filter order 500 | [53] |
| 7 | Denoise | Windowed FIR<br>• 45Hz low pass filter<br>• Hann windows with 50% overlap<br>• Filter order 100 | [53] |
| 8 | Independent component extraction (ICA) | Adaptive ICA (AMICA EEGLab Plugin)<br>• Maximum iterations: 2000<br>• Initial learning rate: 0.1 | [34,60] |
| 9 | Paroxysmal (high-amplitude, non-stationary) artefact removal | Artefact Subspace Reconstruction (ASR)<br>• Burst criterion: 12<br>• Window criterion: 0.15 | [54,69] |
| 10 | Frequency power extraction | Welch PSD computation (pwelch method)<br>• Segment length: 256 samples<br>• Overlap of 64 samples<br>• Hamming windowed<br>• Windows mean averaged<br>• Log-normalization (10*log10) | [22] |
| *Processing inspection & final feature aggregation (R)* | | | |
| 11 | Visual signal inspection for data integrity assessment | • Looking for abnormal channels and high-amplitude artefacts<br>• Inspecting PSD distributions for typical patterns (lower power in higher frequencies) and abnormalities (line noise components or high power in higher frequencies indicating a large amount of muscle artefacts) | [69] |
| 12 | Frequency band aggregation | • Median aggregation used<br>• 4-7 Hz Theta, 8-12 Hz Alpha, 13-30 Hz Beta, 31-40 Hz Gamma | [20,55,69] |

**A.2  System Assembly Instructions**

The herein described system is comprised of three main components: (1) The OpenBCI Cyton microcontroller with the Daisy shield that enables the low-cost mobile biosignal acquisition (e.g. EEG, ECG, or EMG), (2) the cEEGrid electrodes, a set of flexible printed Ag/AgCl electrodes in a c-shaped form that can be placed around a person's ear using adhesives, and (3) the printed circuit board that transmits the signal from the applied cEEGrids to an amplifier. Table 2 lists the required materials together with manufacturers that offer them to enable the assembly of the system. To assemble the system, please follow these instructions:

(1) For the OpenBCI Cyton+Daisy assembly for regular EEG data collection, please follow the thorough instructions provided by the device manufacturers: https://docs.openbci.com/docs/04AddOns/01-Headwear/MarkIV (Last accessed: September 13, 2020). In this documentation, the 3D printed parts are also provided. For the herein presented system to work, the regular board holder (or board mount) needs to be 3D-printed. Also, two #4 screws for brittle plastic are required to attach the clip that is mentioned in step (3). As a power supply, use a ~500mAh lithium-ion rechargeable battery pack that fits into the board holder.

(2) For the assembly of the cEEGrid adapter, we followed the instructions provided by the cEEGrid developers: http://ceegrid.com/home/how-to-connect/ (Last accessed: September 13, 2020). To assemble the adapter, three parts are required, a contact point that connects to the cEEGrid pins (2mm pitch – used here is a mini edge card socket by the company SAMTEC), a simple printed circuit board, and a set of male or female pin headers with 2.54mm pitch (double row). The parts can be joined using a soldering iron. To facilitate the soldering of the mini edge card socket to the PCB and to lower the risk of bridges it is recommended that every second pin (but starting with the first pin) of the card socket is removed before assembling the connector.

(3) Finally, to integrate the two initial components, a few additional steps and components need to be completed. For the herein shown system, a decision was made to solder on stack headers on to the PCB. Initially, individual jumper cables were soldered on to the PCB, yet were found to be a bit cumbersome to work with since not all of the cables can be used with the Cyton+Daisy boards (20 cables for 18 pins). Also, the stack headers provided more flexibility to adapt cable lengths for different mounting solutions. To link the cEEGrid PCB and the OpenBCI board, now only a set of 20 short female jumper cables is required. For this final assembly step, two important aspects need to be mentioned. First, the cables should be attached securely to reduce artefacts that might appear due to cable movement. This factor has been highlighted as a critical issue in related cEEGrid work [11]. As can be seen in Figure 2, we have opted for a simple solution of twisting the cables together to secure them. Second, care needs to be taken to route the cables to the recording pins on the OpenBCI board correctly. While the cEEGrid is symmetrical, the fact that two electrodes need to be left out in this configuration requires an adequate mapping for the left and right ear. To facilitate this step, the schematic in Figure 3 shows the pins on the cEEGrid for the left and right ear. We recommend connecting the right ear to the Cyton pins (channel 1-8 in the OpenBCI GUI) and the left ear to the Daisy pins (channel 9-16 in the OpenBCI GUI). We also recommend maintaining the colour coding shown in Figure 3 to keep track of which electrodes are being used and which are left out. For the correct routing of the reference and ground electrodes, please refer to the OpenBCI documentation mentioned in step (1) or consider the placement in Figure 3 (grey cables). To complete the system, the two parts of the headband clip, the PCB holder, and the re-designed Cyton board cover need to be 3D-printed. The print files are attached as supplementary files. Regular FDM printing can be used with a standard 0.4 mm nozzle diameter and 0.2 mm layer height. Rather slow print speeds (e.g. 40 mm/s) should be used as the parts have fine details. The clip can then be secured on the board holder by using the two #4 screws. The PCB holder can be attached without any additional material. First, the Daisy module should be lifted up. Then the PCB holder can be put in and clipped between the board cover and the Daisy module.

Table 2: Bill of Materials

| Amount | Part Description | Instance / Reference |
|---|---|---|
| *Available Parts* | | |
| 1 | OpenBCI Cyton Board + Daisy Shield (Biosignal Acquisition Boards) | Available from the manufacturer at: https://shop.openbci.com/collections/frontpage/products/cyton-daisy-biosensing-boards-16-channel?variant=38959256526 |
| 2 | cEEGrid Electrodes | Available from the manufacturer at: https://shop.tmsi.com/product/ceegrid |
| *Connector Components* | | |
| 2 | Printed Circuit Boards | Custom PCB made available by the cEEGrid developers here: http://ceegrid.com/download/smarting%20mit%20logo.fzz  Can be manufactured on demand, e.g. at https://aisler.net/ |
| 1 | Re-designed Cyton + Daisy Board Mounts | 3D-Models for 3D-Printing – Attached in Supplementary Files; Board Mounts are adapted from https://github.com/openbci-archive/Docs/tree/master/assets/MarkIV/STL_Directory |
| 2 | PCB Holder | |
| 1 | Headband Clip | |
| 2 | No. 4 Screws for Brittle Plastic | To attach the headband clip to the Cyton board mount - https://www.mcmaster.com/90385A323/ |
| 2 | Mini Edge Card Socket | SAMTEC MB1-120-01-L-S-01-SL-N – configurable at https://www.samtec.com/products/mb1 |
| 20 | Pin Headers | Here we used short male headers with 0.1 inch pitch, e.g. available at https://www.adafruit.com/product/3009 |
| 20 | Jumper Cables | Here we used short (3 inch) female/female jumper wires, e.g. available at https://www.adafruit.com/product/1951 |